\documentstyle[12pt]{article}
\newcommand{\newc}{\newcommand}
\newc{\ra}{\rightarrow}
\newc{\lra}{\leftrightarrow}
\newc{\beq}{\begin{equation}}
\newc{\eeq}{\end{equation}}
\newc{\barr}{\begin{eqnarray}}
\newc{\earr}{\end{eqnarray}}
\def\mue{({\mu^-},{e^-})}

\def\be{\beta}
\def\dl{\delta}

 1

\begin{document} 
\begin{titlepage}

\begin{center}
{\large \bf Muon number violating processes in nuclei}
       
\vspace{12mm}

T.S. KOSMAS$^{a,b}$, J.D. VERGADOS$^a$, and AMAND FAESSLER$^b$

\vspace{2mm}

{a \it Theoretical Physics Division, University of Ioannina,
GR-45110 Ioannina, Greece }

\vspace{2mm}

{b  \it Institut f\"ur Theoretische Physik der Universit\"at
T\"ubingen, D-72076 T\"ubingen, Germany}
\end{center}

\vspace{7mm}

\begin{abstract}
{\it The flavour violating neutrinoless muon decays in the presence of nuclei,
are discussed. We focus on the theoretical aspects of 
$\mu^-(A,Z)\to e^-(A,Z)^*$ (muon to electron conversion), one of the most
prominent flavour changing reactions, emphasizing its connection with the
physics beyond the standard model. This process offers the most severe limits 
of the lepton flavour violation. By using the nuclear transition matrix 
elements calculated with several methods, and the recent experimental data of
the branching ratio $R_{\mu e^-}$, we determine limits for the flavour
changing parameters entering the elementary sector part of $R_{\mu e^-}$. 
These results are discussed in view of the ongoing experiment at PSI and the
designed at Brookhaven, which are expected to push down by some orders of 
magnitude the experimental sensitivity the next few years with the hope to
see ``new physics''. }
\end{abstract}

\vspace{1.0cm}

\end{titlepage}
\section{ Introduction}

In the framework of the standard model of the electroweak interactions, if one 
associates with each lepton generation a lepton flavour quantum number as
\beq
\pmatrix{e^- \cr \nu_e \cr} \to L_e, \quad \pmatrix{\mu^- \cr \nu_{\mu} \cr} 
\to L_{\mu}, \qquad \pmatrix{\tau^- \cr \nu_{\tau} \cr} \to L_{\tau}
\label{eq:1.1}
\eeq
(for antiparticles $L_e$, $L_{\mu}$, $L_{\tau}$ take values with opposite
sign), the quantum numbers $L_e, \,\, L_{\mu}, \,\, L_{\tau}$ 
are separately conserved. In addition, the total lepton number
\beq
L \, = \, L_e + L_{\mu} + L_{\tau}
\label{eq:1.2}
\eeq
appears also to be conserved~\cite{Engf}-\cite{Depom}. Even though these
conservation
laws are uncontradicted with experiment~\cite{Walt,Honec}, the standard
model is believed to be an effective low
energy approximation of a more fundamental theory which is unbroken at
high energies. In this spirit, many theories beyond the standard model
have been developed during the last decades (grand unified models, 
supersymmetric theories etc.)~\cite{WeiFei}-\cite{KLV89}.
in which the existence of flavour and lepton changing effective currents
is predicted. In this way, a plethora of
exotic processes, non-conserving the flavour, $L_e , L_{\mu} , L_{\tau}$, 
and/or lepton, $L$, quantum numbers, are possible. For some examples
and the existing limits see Table 1 \cite{JDV,WeiFei}-\cite{muplus}.

In the present work we are going to discuss the category of 
neutrinoless exotic muon decays which take place in a muonic atom, 
i.e. in the case when a muon is bound in the field of a nucleus.
In the context of the aforementioned
extensions of the standard model this bound muon 
can give two yet non-observed processes; the 
conversion of $\mu^-_b$ into an electron~\cite{Stein}-\cite{SFK97},
\begin{equation}
 \mu^-_b \, + \, (A,Z)\,\, \to \,\, e^- \, + \, (A,Z)^{\ast}
 \qquad \qquad (\mu^-_b \to e^- \, conversion)
\label{1.3}
\end{equation}
which violates the lepton flavour quantum numbers $L_e$ and $L_{\mu}$, and the
conversion of the muon $\mu^-_b$ into a positron~\cite{muplus},
\begin{equation}
 \mu^-_b \, + \, (A,Z) \,\, \to \,\, e^+ \, + \, (A,Z-2) \qquad \qquad
(\mu^-_b \to e^+ \,\, conversion)
\label{1.4}
\end{equation}
which violates the conservation of the total lepton $L$ as well as
the flavour $L_e$ and $L_{\mu}$ quantum numbers.

The important role played by the muon as a test particle in the investigation 
of the electromagnetic and weak interactions with hadronic particles,
has been recognized decades ago~\cite{Scheck}.
The bound in the innermost 1s atomic orbital muon $\mu^-_b$ 
has especially prominent role, since it 
gives rise to very important allowed decay modes~\cite{Goul-Prim,Suzu}.
it can disappear either by decaying into an electron and two neutrinos,
\begin{equation}
\mu^-_b \to e^- + \bar{\nu}_e + \nu_\mu \qquad (muon \,\, decay \,\,
in \,\, orbit),
\label{1.5}
\end{equation}
or by being captured by the nucleus~\cite{Goul-Prim,Suzu},
\begin{equation}
\mu^-_b + (A,Z) \to \nu_\mu + (A,Z-1) \qquad (ordinary \,\, muon \,\, capture)
\label{1.6}
\end{equation}
The latter two processes conserve both lepton and flavour 
quantum numbers and have been well studied 
theoretically as well as experimentally~\cite{Goul-Prim,Suzu}.

The exotic channels of Eqs. (\ref{1.3}) and (\ref{1.4}),
are the subject of some of the most
important recent experiments~\cite{Walt,Honec} 
searching extensively for events involving muon number non-conservation.
Up to now, these experiments have only put upper limits on the branching 
ratios $R_{\mu e^-}$ and $R_{\mu e^+}$, i.e. on the rates
of processes (\ref{1.3}) and (\ref{1.4})
divided by the total rate of the 
ordinary muon capture reaction.
For a comparison of these upper limits with those obtained from
other lepton flavour violating reactions see Table 1. 
Obviously, the observation of any of these processes will signal ``new physics''
beyond the standard model and restrict the various models predicting such 
exotic reactions.

\subsection{Basic features of the exotic muon decays in nuclei}

The neutrinoless rare decays,
$\mu^-_b \to e^-$ and $\mu^-_b \to e^+$ conversions in nuclei, start from the
same initial state (a muon at rest in a muonic atom). Experimentally, 
they are both simultaneously studied. Theoretically,
both of them can occur by assuming mixing of intermediate (massive) 
neutrinos, but $(\mu^-,e^-)$ cannot distinguish between Dirac and Majorana
neutrinos while
$(\mu^-,e^+)$ can proceed only if the neutrinos are Majorana particles.
These processes can also be predicted by supersymmetry via the mixing 
of the intermediate supersymmetric particles, e.g. s-leptons
(see sect. 3). 

There are, however, big differences between the exotic decays $(\mu^-,e^-)$
and $(\mu^-,e^+)$ originating from the 
individual characteristics of their transition operators. Thus, due to its
charge conserving operator, the $\mu^-_b \to e^-$ conversion can take
place in one nucleon; it can proceed via the coherent mode (when the
participating nucleus remains in its ground state and all nucleons
participate)~\cite{Shank}, the only measurable $(\mu^-,e^-)$ conversion
channel, which is an important experimental feature.

Another experimental advantage of the $\mue$ we should mention is 
the fact that, a
detection of only one particle is sufficient and there is no need for
coincidence measurement~\cite{Walt}. If the electron energy is equal to
$E_e \approx m_{\mu} c^2 -\epsilon_b$ (coherent mode) the process is 
free from the reaction induced background which involves,
in addition to the muon decay in orbit mentioned above, 
the following reactions:

i) Muon decay in flight. 
\begin{equation}
 \mu^- \, \rightarrow  \, \nu_\mu  \, + 
{\bar \nu}_e \, + e^-
\label{1.7}
\end{equation}

ii) Radiative muon capture followed by the creation of a fully asymmetric
$e^-, \,\,\, e^+$ pair at zero energy, i.e.

\begin{eqnarray}
\mu^-_b \, + \,(A,Z) \rightarrow (A,Z-1)\,\, + \,\, \nu_\mu  \, + \,\,
&&{\gamma}  \label{1.8} \\
&&^{\mid}\to  e^+ \, +\, e^- 
\nonumber
\end{eqnarray}

On the other hand, the muon to positron conversion
$(\mu^-,e^+)$, is more complex since it is a
two-nucleon process; it is similar to but more complicated than
the neutrinoless double beta decay
\begin{equation}
(A,Z) \rightarrow (A,Z{\pm}2) + e^{\mp} e^{\mp} \qquad \qquad
(0\nu  \beta \beta -decay) 
\label{1.9}
\end{equation}
which is a lepton and flavour number violating process
extensively discussed in the present workshop (see e.g.~\cite{Koval}).

Because of its simplicity and important characteristics,
the $(\mu^-,e^-)$ conversion is preferentially studied by many
authors~\cite{WeiFei}-\cite{KV96}, and
in the present work we will devote to it a rather extensive discussion.
We pay special attention to the construction of
the effective Hamiltonian in the framework of several common extensions of the
standard model. This is subsequently used to calculate the relevant 
transition rates. Using the results of the branching ratio $R_{\mu e^-}$ and
exploiting the existing experimental data we put limits on the flavour 
violating parameters (see sect. 6).

\section{The $\mue$ conversion in Nuclei}

\subsection{ Brief historical review}

The first experimental searching
for $\mue$ events has been done long ago by Steinberger 
and Wolfe~\cite{Stein} at Columbia Cyclotron using $Cu$ as a target. This 
experiment put an upper limit for the branching ratio $R_{\mu e^-}$ 
\beq
R_{\mu e^-}\, =\, \frac{\Gamma(\mu^- \ra e^-)}{\Gamma(\mu^- \ra capture)}
\label{2.1.1}
\eeq
on the value $R_{\mu e^-} \, < \, 5 \times10^{-4}$. Some years
later Conversi {\it et al.}~\cite{Conver} in the synchrocyclotron at CERN
using again $Cu$ as target with two independent experiments found the limits
$ R_{\mu e^-} \, < \, 5 \times10^{-5}$ and 
$R_{\mu e^-} \, < \, 5 \times10^{-6}$.
A $Cu$ target was also used at TRIUMF by Bryman {\it  et al.}~\cite{Brym72}
in the TPC experiment. From the data of this experiment
the limit $ R_{\mu e^-} \, < \, 2.6 \times10^{-8}$ was extracted.

We should note that, the use of a $Cu$ target in the aforementioned 
experiments had been influenced by the theoretical estimates of Weinberg
and Feinberg~\cite{WeiFei} who found
that, the $\mu^-_b \to e^-$ conversion branching ratio shows a maximum 
in the region of $Cu$ (see also Ref.~\cite{KV88}).

Target different than $Cu$ was first used at SIN (present PSI) by Badertscher
{\it  et al.}~\cite{Bader82} in the SINDRUM spectrometer who used
the isoscalar target $S$. The upper limit set on $R_{\mu e^-}$ 
by this experiment was $R_{\mu e^-} \, < \, 1.6 \times10^{-8}$.

In another TPC experiment at TRIUMF (Bryman {\it et al.}~\cite{Brym85})
a $Ti$ target was used and pushed the upper limit of
$R_{\mu e^-}$ at first to the value $R_{\mu e^-}\,<\,1.6 \times10^{-11}$
and some years later~\cite{Ahmad} to the value
$R_{\mu e^-}\,<\,4.6 \times10^{-12}$.

The best upper limit on $R_{\mu e^-}$ set up to the present has been
obtained by using $Ti$ target at PSI~\cite{Walt} which yielded the value
$ R_{\mu e^-} \, < \, 7.0 \times10^{-13}$. Recently, by using
a heavy $^{208}Pb$ target in the SINDRUM II experiment the limit
$R_{\mu e^-} \, < \, 4.6 \times10^{-11}$ was determined~\cite{Honec}.
This limit was an improvement by an order of magnitude over
the previous limit $R_{\mu e^-} \, < \, 4.9 \times10^{-10}$ obtained from
preliminary results on $Pb$ at TRIUMF (Ahmad {\it et al.}~\cite{Ahmad}). 

Ongoing experiments at PSI as well as planned experiments at Brookhaven, 
discussed by Walter~\cite{Walt} in this conference, 
are expected to increase the
experimental sensitivity by two to three orders of magnitude with
the hope to see some events of the reaction (\ref{1.3}) in the near future
or to drastically push down the best upper limit of the branching 
ratio $R_{\mu e^-}$.

From a theoretical point of view, the basic background for the reaction
$\mu^-_b \to e^-$ has been formulated by 
Weinberg and Feinberg~\cite{WeiFei} who assumed that this process is
mediated by a virtual photon (photonic mechanism). 
Non-photonic contributions
were included later on in the post-gauge-theory 
era~\cite{MarSan}-\cite{Berna}.
Afterwards, a lot of new mechanisms which lead to the
$\mu - e $ conversion have been proposed~\cite{KFV97}.
The mechanisms which involve intermediate neutrino mixing or
mixing of intermediate s-fermions  
are extensively discussed in the next section.

From a nuclear physics point of view, it is important to know 
the transition matrix elements
of the participating nucleus in processes (\ref{1.3}) and (\ref{1.4}).
For the reaction (\ref{1.3}) this was recently done for a series of nuclear 
isotopes~\cite{KFSV97} including $^{48}Ti$ and $^{208}Pb$ 
which are of current experimental interest~\cite{Honec,Dohm}.
Several models, like shell model, local density
approximation, quasi-particle RPA etc., have been employed (see sect. 4).

\section{ Effective Hamiltonian of $\mue$ in common extensions of SM}

Some Feynman diagrams, which contribute to $\mu^-_b \to e^-$ conversion
in the lowest order of perturbation theory (one loop level) are shown in
Figs. 1 and 2. In Fig. 1 mixing of intermediate neutrinos and gauge bosons 
is involved and in Fig. 2 the mixing of neutralinos and s-leptons in a
supersymmetric model is considered.
In the context of these models we construct below the effective Hamiltonian
for the $(\mu^-,e^-)$ conversion by first writing down the hadronic and
leptonic currents of these diagrams as follows.

\subsection{ Hadronic currents at the nucleon level}

Usually one starts from the weak vector and axial vector quark currents. 
By assuming a specific nucleon model, one transforms these hadronic currents to 
the nucleon level. For the diagrams shown in Figs. 1 and 2,
one gets the following results.

{\it i) Photonic diagrams} (Figs. 1(a) and 2(a)): In this case the hadronic
vertex is the usual electromagnetic coupling
\begin{equation}
J^{(1)}_{\lambda} = {\bar N}_p \gamma_{\lambda} N_p 
 \,= \,{\bar N} \gamma_{\lambda} \frac{1}{2}\,(1 + \tau_3)\, N 
\qquad (photonic )
\label{3.1.3}
\end{equation}
$N$ is the nucleon isodoublet, $N= \pmatrix{ N_p \cr N_n \cr}$, with
$N_p$, $N_n$ the proton, neutron spinors.

{\it ii) Non-photonic diagrams}: There are many such diagrams contributing
to the $\mu^-_b \to e^-$ conversion.
The most obvious are:
a) those mediated by Z-particle exchange (see Figs. 1(a),(b), 2(a),(b)) and 
b) the box diagrams involving either
massive neutrinos and W-bosons, as in Fig. 1(c), 
or s-leptons and neutralinos, as in Fig. 2(c).
The non-photonic hadronic current which includes all the above cases 
can be compactly written as
\begin{equation}
J_{\lambda}^{(2)} = {\bar N}\gamma_{\lambda} \frac{1}{2}\left[
(3 + f_V \beta \tau_3) - (f_V\beta^{''}+f_A \beta^{'} \tau_3)
\gamma_5 \right ] N \qquad (non-photonic)
\label{3.1.4}
\end{equation}
where the parameters $\beta$, $\beta^{\prime}$, $\beta^{\prime \prime}$,
for the models assumed take the values
\begin{equation}
\beta  \, = \,
\left\{ \begin{array}{ l@{\quad \quad} l}
5/6,& $ $ W-boson $ $ exchange $ $ $ $   \\
3/5,  & $ $ SUSY $ $ Box $ $ diagrams $ $ $ $  \\ 
-3.46,  & $ $ SUSY $ $Z-exchange $ $ $ $  \\ 
\end{array} \right. 
\label{3.1.5}
\end{equation}
\begin{equation}
\beta^{\prime} \, = \,
\left\{ \begin{array}{ l@{\quad \quad} l}
\beta,    & $ $ for $ $ box $ $ diagrams $ $ $ $  \\
6.90,  & $ $ $ $Z-exchange $ $ $ $  \\ 
\end{array} \right. 
\label{3.1.6}
\end{equation}
\begin{equation}
\beta^{\prime \prime} \, = \,
\left\{ \begin{array}{ l@{\quad \quad} l}
0,  & $ $ SUSY $ $Z-exchange $ $ $ $ \\ 
1,    & $ $ all $ $ other $ $ cases $ $ $ $  \\
\end{array} \right. 
\label{3.1.7}
\end{equation}
The parameter $\beta$ is defined as $\beta =\beta_1/\beta_0$, with 
$\beta_0$ $(\beta_1)$ the isoscalar (isovector) couplings at the quark
level and $f_V$, $f_A$ represent
the vector, axial vector static nucleon form factors ($f_A/f_V = 1.24$).
We should note that, in general, the parameters $\beta$, $ \beta^{'}$ and
$\beta^{''}$ are functions of $sin^2\theta_W$.
For example, in the case of Z-exchange
we have $\beta^{'}=3/2sin^2\theta_W=6.90$ (for some other cases see 
Ref.~\cite{KFV97}). 

\subsection{The leptonic currents}

The corresponding leptonic currents in the diagrams of Figs. 1 and 2 
are given from the following expressions.

{\it i) photonic mechanism:}

\begin{equation}
j^{\lambda}_{(1)} = {\bar u}(p_e) \Big[ \, (f_{M1} + \gamma_5 f_{E1}) i
\sigma^{\lambda \nu} \frac{q_{\nu}}{m_{\mu}} +
\left (f_{E0} + \gamma_5 f_{M0}\right) \gamma_{\nu}  \left (
g^{\lambda \nu} - \frac {q^{\lambda} q^{\nu}}{q^2}\right )
\Big] u(p_{\mu})
\label{3.2.1}
\end{equation}

{\it ii) Non-photonic mechanism:}

\begin{equation}
j^{\lambda}_{(2)} = {\bar u}(p_e)\gamma^{\lambda}
\, {\frac{1}{2}} \, (f_1 + f_2 \gamma_5) u(p_{\mu})
\label{3.2.2}
\end{equation}

\noindent
where $q=p_e - p_{\mu}$ is the momentum transfer with
$p_e,\,\,p_\mu$ the lepton momenta. The parameters
$f_{E0}$, $f_{E1}$, $f_{M0}$, $f_{M1}$, $f_1$, $f_2$ are
gauge model dependent form factors. In the models mentioned above 
they are interelated as follows.

\subsubsection{Models involving intermediate neutrinos and gauge bosons}

For purely left-handed theories we have
\begin{equation}
f_{E0} = -f_{M0}, \qquad f_{E1} = -f_{M1}, \qquad f_1  = -  f_2
\nonumber
\label{3.2.3}
\end{equation}
In models involving $W$-bosons, it, furthermore, holds 
\begin{equation}
\frac {f_{E0}}{q^2} = - \frac{f_{E1}}{m^2_\mu}
\label{3.2.4}
\end{equation}

\subsubsection{Models involving intermediate neutralinos and s-leptons}

In this case, the photonic and non-photonic parameters are related as
\begin{equation}
(4\pi\alpha)f_{M1}=-(4\pi\alpha)f_{E1}=-\frac{1}{24}f 
\label{3.2.5}
\end{equation}
\begin{equation}
(4\pi\alpha)f_{E0}=-(4\pi\alpha)f_{M0}=-\frac{1}{72}f 
\label{3.2.6}
\end{equation}
\begin{equation}
f_{1}  =  -f_{2}=\frac{1}{16}\beta_0 f, \qquad \quad
\beta_0=-\frac{2}{3} sin^2  \theta_w
\label{3.2.7}
\end{equation}
\begin{equation}
f  =  \alpha^{2}\frac{m^2_{\mu}}{\bar{m}^2}\tilde\eta, \qquad\qquad 
\tilde\eta  =  \frac{\left(\delta m^2_{ll}\right)_{12}}{\bar{m}^2} 
\label{3.2.8}
\end{equation}
$\bar m^2$ are the average of the square of the s-fermion masses
and $\bar \eta$ depends on the details of the model. 
For Z-exchange contribution in SUSY models
(see also Ref.~\cite{KFV97}) we have

\begin{equation}
f_{1}  =  -f_{2}\, =\, \xi \alpha^{2} \frac{m^2_{\mu}}{m^2_Z} \tilde\eta
\label{3.2.9}
\end{equation}
$\xi$ depends on the specific model; it vanishes
if one ignores the Higgsino components of the neutralinos (e.g. in the case
of a pure photino); it also vanishes if the  probabilities of finding the two
Higgsinos in the neutralino are equal. In the model 2 of Ref.~\cite{KVprd}
it takes the value $\xi =2.8\times 10^{-2}$. In models in which $\bar{m}^2$ 
is larger than $m^2_Z$, the Z-exchange is more favored. Due to the
smallness of $\xi$, however, many authors consider $Z'$ etc.~\cite{Berna}.

\subsection{The $\mue$ conversion effective Lagrangian at nucleon level}

The effective Lagrangian at nucleon level, which contains contributions
from all types of diagrams mentioned above, is written in the form 
\begin{equation}
{\cal M} \,=\,\frac{4\pi\alpha}{q^2} \, j^{\lambda}_{(1)} J^{(1)}_{\lambda}
\, + \, \frac{\zeta}{m^2_{\mu}} \, j^{\lambda}_{(2)} J^{(2)}_{\lambda}
\label{3.3.1}
\end{equation}

\hspace{3.75cm}${ \it (Photonic) \qquad  (Non-photonic)} $

\noindent
where $\alpha$ is the fine structure constant, and
\begin{equation}
\zeta \,= \, 
\left\{ \begin{array}{ l@{\quad \quad} l}
\frac{ G_Fm^2_{\mu}}{\sqrt 2}, \qquad & W-boson \,\,\, exchange \\
1,  & $ $ $ $ s-lepton $ $ $ $ mixing$ $ $ $  \\ 
\end{array} \right. 
\label{3.3.2}
\end{equation}

\subsection{Transition operators at the nuclear level}

In order to obtain the transition operators at the nuclear level, 
where the calculations of the transition rates are done, one 
usually assumes that the lower components of the nucleon spinors is negligible.
This assumption gives the non-relativistic limit of hadronic currents.
For a nucleus $(A,Z)$ this leads to the following $\mu^-_b \to e^-$ 
conversion operators

\begin{equation}
\Omega_0 = \tilde g_V{\sum^{A}_{j=1}}\Big (
 {3+f_V\be\tau_{3j}}\Big)
e^{-i{\bf q} \cdot {\bf r}_j} 
\qquad \qquad \qquad (Vector)
\label{3.4.1}
\end{equation}
which is of Fermi-type, and
\begin{equation}
{\bf \Omega} =-\tilde g_A {\sum^{A}_{j=1}}\Big (
{f_V\beta '' +f_A \be '\tau_{3j}}\Big) {{{\bf \sigma}_j}\over
\sqrt 3}\,e^{-i{\bf q} \cdot {\bf r}_j} 
\qquad \qquad (Axial \quad Vector)
\label{3.4.2}
\end{equation}
which is of Gammow-Teller type ($f_V=1.0, \,\, f_A=1.24$). The parameters
${\tilde g}_V$ and ${\tilde g}_A$ take the values
${\tilde g}_V= 1/6, \quad {\tilde g}_A = 0$, 
for the photonic case, and ${\tilde g}_V={\tilde g}_A= 1/2$,
for the non-photonic case.

In the calculations of the nuclear part of the transition rate, the following
multipole expansion operators, obtained from Eqs. (\ref{3.4.1}),
(\ref{3.4.2}), enter.

\begin{equation}
{\hat T}_M^{(l,0)J}= \dl_{lJ} \, \sqrt{4\pi}\, {\sum_{i=1}^A}
(3+  f_V \be \tau_{3i} )
j_l (kr_i)Y_M^l ({\bf {\hat r}}_i)  
\label{3.4.3}
\end{equation}
\begin{equation}
{\hat T}_M^{(l,1)J}=\,\sqrt{4\pi}\,\, {\sum_{i=1}^A} (f_V \beta '' +
f_A \be ' \tau_{3i}) j_l(kr_i)
\Big[ {Y^l({\bf {\hat r}}_i) {\bf \times} {\bf \sigma}_i} \Big] _M^J
\label{3.4.4}
\end{equation}
The matrix elements of these operators calculated in a given nuclear model 
provide the individual multipole contributions 
(see Refs.~\cite{KVCF,KFSV97,SFK97}).
From kinematics one can show that the momentum transfer q can be approximated
by
\begin{equation}
q \,\equiv q_f =\, m_\mu - \epsilon_b -(E_f - E_{gs})
\label{3.4.5}
\end{equation}
where $\epsilon_b$ represent the muon binding energy and
$(E_f-E_{gs})$ the excitation energy of the nucleus (A,Z).

\section{\large \bf The $\mue$ conversion branching ratio $R_{\mu e^-}$}

In studying the $\mu^- \to e^-$ conversion process, the most interesting
quantities both theoretically and experimentally are (i) the
branching ratio $R_{\mu e^-}$ defined in Eq. (\ref{2.1.1})
and (ii) the ratio of coherent to total rate
\begin{equation}
\eta \, = \,  \frac{\Gamma_{coh}(\mu \ra e^-) }
{\Gamma_{tot}(\mu \ra e^-) }
\label{4.1}
\end{equation}

In general, it is impossible to
separate the nuclear structure dependence of $R_{\mu e^-}$ from the 
elementary particle parameters. In the coherent mode, however,
this is sometimes possible (see Ref.~\cite{KFV97}).
In the models mentioned in the previous section the 
dominant coherent channel, can be written as
\begin{equation}
R_{\mu e^-} = \rho \, \gamma
\label{4.2}
\end{equation}
where $\rho$ is independent on nuclear physics. The function $\gamma(A,Z)$ 
contains all the nuclear information and is defined as
\begin{equation}
\gamma \, = \, \frac{E_e p_e}{m_\mu^2} \,
\frac{ M^2}{G^2 Z f_{GP}(A,Z)}
\label{4.3}
\end{equation}
($G^2 \approx 6$) with $f_{GP}$ the Goulard-Primakoff~\cite{Goul-Prim}
function describing the total $\mu^-_b \to \nu_\mu$ rate and
$M^2$ the nuclear transition matrix elements of the $\mu^-_b \to e^-$
conversion. Thus,
the nuclear aspect of the $\mu-e$ conversion process involves
the matrix elements $M^2$ which in general are given by
\begin{equation}
M^2 \,\, =\,\,
f_V^2 \,\mid \langle f \mid \Omega_0 \mid i,\mu\rangle \mid ^2
+3 f_A^2 \,\mid \langle f \mid {\bf \Omega} \mid i,\mu\rangle \mid ^2
\label{4.4}
\end{equation}
where $| f \rangle$ the final nuclear state populated during the process.

The quantity $\eta$ of Eq. (\ref{4.1}), in the case of the approximation Eq.
(\ref{4.2}) can be written in terms of the matrix elements of Eq. 
(\ref{4.4}) as
\begin{equation}
\eta \, = \, \frac{\Gamma_{coh}(\mu \ra e^-)}{\Gamma_{tot}(\mu \ra e^-)}
\, \approx \, \frac{M^2_{coh}}{M^2_{tot}}
\label{4.5}
\end{equation}
where $M^2_{coh}$, $M^2_{tot}$ the coherent, total rate matrix elements,
respectively.  Such calculations of $\eta$ are discussed in sect. 6.

\subsection{Definitions of $\gamma(A,Z)$  and $\rho$ in various models}

{\it i) Neutrino mixing models}: For the photonic diagrams, the
function $\gamma$ of Eq. (\ref{4.3}) takes the form
\begin{equation}
\gamma_{ph} \,=\, \frac{E_e p_e}{m_\mu^2}\, 
\frac{Z |F_Z(q^2)|^2 }{G^2 f_{GP}(A,Z)}
\label{4.1.1}
\end{equation}
For the photonic mechanism with left-handed currents only, the parameter
$\rho$ of Eq. (\ref{4.2}) is given by
\begin{equation}
\rho = (4\pi\alpha)^2\frac{\vert f_{M1}+f_{E0}\vert ^2+  
\vert f_{E1}+f_{M0}\vert^2}{(G_Fm^2_{\mu})^2}
 = \frac{9\alpha^2}{64\pi^2}\vert\frac{m^2_e}{m^2_W}\eta_{\nu}+\eta_N 
    \vert ^2 
\label{4.1.2}
\end{equation}
For the non~-~photonic diagrams the function $\gamma$ is given by Eq. 
(\ref{4.3}) with $M^2$ given by
\beq
M^2 =  \left [1+\frac{3-f_V\beta}{3+f_V\beta}\,
\frac{N}{Z} \, \frac{F_N(q^2)}{F_Z(q^2)}\right ]^2 \gamma_{ph}
\label{4.1.3}
\eeq
($F_Z(q^2)$, $F_N(q^2)$ are the proton, neutron nuclear form factors)
and the parameter $\rho$ associated with intermediate neutrinos is
given by
\begin{equation}
\rho = \frac{\vert\beta_0f_1\vert^2+\vert\beta_0f_2\vert^2}{2}
     = \frac{9}{64\pi^4}(G_Fm^2_W)^2\vert 30\frac{m^2_e}{m^2_W}\eta_{\nu}
	 +\eta_N\vert ^2
\label{4.1.4}
\end{equation}

The quantities $\eta_{\nu}$ and $\eta_N$ in Eqs. (\ref{4.1.2}) and 
(\ref{4.1.4})
are the flavour violating parameters associated with intermediate light 
$(\nu_j)$ or heavy $(N_j)$ neutrinos. They depend on the gauge model 
(we have ignored the neutrino mass independent contribution arising
from the non-unitarity of $U^{(1)}$) as
\begin{equation}
\eta_\nu = \sum_j U^{(1)}_{ej} U^{(1)*}_{\mu j} \frac{m^2_j}{m^2_e}
\label{4.1.5}
\end{equation}
\begin{equation}
 \eta_N = \sum_j U^{(2)}_{ej} U^{(2)*}_{\mu j} \frac{m^2_W}{M^2_j}
\left( -2 ln \frac{ M^2_j}{m^2_W} +3 \right) 
\label{4.1.6}
\end{equation}
$U^{(i)}_{ej}$, $U^{(i)}_{\mu j}$, with $i=1$ for light neutrinos
and $i=2$ for heavy neutrinos, are the 
elements of the charged lepton current mixing matrix 
associated with the electron and the muon as
\begin{equation}
\nu_e =\sum_{j=1}^{n} U^{(1)}_{ej} \nu_j+\sum_{j=1}^{n}U^{(2)}_{ej} N_j
\label{4.1.7}
\end{equation}
\begin{equation}
\nu_{\mu} =\sum_{j=1}^{n} U^{(1)}_{\mu j} \nu_j+     
\sum_{j=1}^{n}U^{(2)}_{\mu j} N_j \label{eq:2.2}
\label{4.1.8}
\end{equation}
where $\nu_e, \, \nu_\mu$ are the weak neutrino eigenstates.
The number $n$ counts the generations assumed. 

{\it ii) Neutralino and s-lepton mixing models}:
Under some plausible assumptions and neglecting the Z-exchange, 
the function $\gamma$ and the parameter $\rho$ are given by 
\begin{equation}
\gamma \,=\, \zeta \,\Big( \frac{13}{12} + \frac{1}{2} 
\,\frac{N}{Z} \, \frac{F_N}{F_Z} \Big)^2 \, \gamma_{ph}
\label{4.1.9}
\end{equation}
\begin{equation}
\rho \,=\,\frac{1}{288}\frac{\alpha^4}{(G_F {\bar m}^2)^2}\,
|{\bar \eta}|^2 
\label{4.1.10}
\end{equation}
In the case of Z-exchange diagrams we have
\begin{equation}
\gamma \,=\, 0.053 \Big(1  - 14.04\,\frac{N}{Z} \, \frac{F_N}{F_Z}
\Big)^2 \gamma_{ph}
\label{4.1.11}
\end{equation}
and the quantity $\rho$ is now given by
\begin{equation}
\rho \,=\,5.5\times 10^{-4}\frac{\alpha^4}{(G_F {\bar m}^2)^2}\,
|{\bar \eta}|^2 
\label{4.1.12}
\end{equation}

We should note that, in the models discussed above,
the only variables of the elementary particle sector which enter
the function $\gamma$ are the parameters
$\beta =\beta_1/\beta_0,\,\, \beta '$ and $\beta ''$. 
Once $\gamma(A,Z)$ is known, e.g. by nuclear model calculations,
from Eq. (\ref{4.2}) one can extract information about the 
interesting parameter $\rho$ from the experimental data
(see sects. 5 and 6).
 
\section{Nuclear model calculations for the branching ratio $R_{\mu e^-}$}

The main aim of the nuclear calculations of the $\mu^- \to e^-$ conversion
is to study the nuclear structure dependence
of its coherent, incoherent and total rates throughout the periodic
table and find favorable nuclear systems to be used as targets
in searching for lepton number non-conservation.
Furthermore, by exploiting the nuclear part of the branching ratio
in Eq. (\ref{4.2}) we can put restrictions on the elementary sector part
of $R_{\mu e^-}$, i.e. the quantity $\rho$,
which contains the lepton flavour violating parameters
(neutrino masses, mixing angles etc.) of a specific gauge model. 
In this way, one can check the models predicting flavour
non-conservation and improve them.
 
\subsection{Calculation of the muon-nucleus overlap integral}

As we have seen in sect. 4, by neglecting the effect of nuclear recoil,
the nuclear dependence of the $\mu-e$ conversion rate is, in general,
included in the matrix elements of Eq. (\ref{4.4}). 

In the special case of the coherent process, i.e. ground state to ground 
state transitions, only the vector component of Eq. (\ref{4.4}) contributes
and one obtains
\begin{equation}
\langle f \mid \Omega_0 \mid i,\mu \rangle \, = \tilde g_V \,(3 + f_V \beta)
 \, \tilde F(q^2)
\label{5.1.1}
\end{equation}
where ${\tilde F}(q^2)$ is the  matrix element involving the ground state 
proton, $\rho_{p}({\bf x})$, and neutron, $\rho_{n}({\bf x})$, 
densities (normalized to Z and N respectively) as
\begin{equation}
\tilde F (q^2) = \tilde F_{p} (q^2) +
\frac{3-f_V\beta}{3+f_V\beta}
\tilde F_{n} (q^2)
\label{5.1.2}
\end{equation}
with
\begin{equation}
\tilde F_{p,n} (q^2) = \int d^{3}x \; \rho_{p,n} ({\bf x}) \; e^{-
i{\bf q} \cdot {\bf x}}\; \Phi_{\mu} ({\bf x})
\label{5.1.3}
\end{equation}
Thus, the coherent rate matrix elements can be obtained from Eq.
(\ref{5.1.1}) by firstly calculating $\tilde F(q^2)$ from Eq.
(\ref{5.1.2}) for a given muon wave function and given proton, neutron
nuclear density distributions needed in Eq. (\ref{5.1.3}).

The methods used up to now, according to how do they treat the muon-nucleus
overlap integral appeared in the branching ratio, are classified in two
categories: i) those using exact evaluation of the muon wave function, and
ii) those using the factorization approximation for the muon wave function.
In the next subsections we present a brief description of these methods.

\subsubsection{Exact Evaluation of the muon wave function}

For point like nuclei the muon wave function $\Phi_{\mu}({\bf r})$
is trivially obtained by solving the Schr\"odinger or Dirac
equations for the Coulombic nuclear potential. By
taking into consideration the effects of finite nuclear size and 
vacuum polarization one needs to solve numerically these equations.
Such calculations were done in Refs.~\cite{Chiang,LagKos}.
In Ref.~\cite{Chiang} the muon wave function was
obtained by solving numerically the Schr\"odinger equation 
taking into account the aforementioned effects.
In Ref.~\cite{LagKos} the Dirac and Schr\"odinger equations 
are solved using modern neural network techniques and an analytic
expression for $\Phi_{\mu} ({\bf r})$ is obtained by means of
optimization methods.

\subsubsection{Factorization approximation for the muon wave function}

A reasonable assumption for the $1s$ muon wave function of
light and medium nuclei is to assume that it
varies little inside such systems. In this case 
one can use for the matrix elements Eq. (\ref{5.1.1}) the following
approximation
\beq
\langle f \mid \Omega_0 \mid i,\mu\rangle \, \approx \, 
{\tilde g}_V (3+f_V\beta) \,
\langle\Phi_{1s} \rangle \langle f \mid \Omega_0 \mid i\rangle 
\label{5.1.4}
\eeq
where $\langle \Phi_{1s} \rangle$ is the mean value of the muon wave function
given by
\begin{equation}
\langle \Phi_{1s}\rangle^{2} = \frac{\alpha^{3} m_{\mu}^{3}}{\pi} \;
\frac{Z_{eff}^{4}}{Z}
\label{5.1.5}
\end{equation}
$\alpha$ is the fine structure constant and
$Z_{eff}$ represents the effective charge which sees the $\mu^-_b$
in a muonic atom. For light nuclei $\mu^-_b$ lives outside
the nucleus, $Z_{eff} \approx Z$, but for heavy and very heavy nuclei
the $\mu^-_b$ resides inside the nucleus and $Z_{eff} \ll Z$.

Using this approximation, for the coherent rate 
many authors (see Ref.~\cite{WeiFei,Shank,KV88})
calculated the quantity ${\tilde F}(q)$ in
eq. (\ref{5.1.2}) from the equation
\begin{equation}
\vert \tilde F (q^2) \vert^{2} \approx
\frac{\alpha^{3} m_{\mu}^{3}}{\pi} \,
\frac{Z_{eff}^{4}}{Z} \, \vert  ZF_Z(q^2)+
\frac{3-f_V\beta}{3+f_V\beta} \, NF_N(q^2) \vert^{2}
\label{5.1.6}
\end{equation}
The nuclear form factors ${F}_{Z}$, ${F}_{N}$ 
are either calculated by using various models, i.e.
shell model~\cite{KV88,KV92}, quasi-particle RPA~\cite{KVCF} etc., or 
extracted from experimental data whenever possible~\cite{Heisen,Vries}.

The branching ratio $R_{\mu e^-}$ in this approximation takes the form
of Eq. (\ref{4.2}) with $\gamma$ given as explained in sect. 4.

\subsection{Evaluation of the partial $\mu^- \to e^-$ conversion rates}
 
The methods used up to now, according to how do they 
evaluate the partial rates, i.e. the transition rate of each individual 
channel, are classified in two categories:
1) those doing state-by state calculations of the partial rates
2) those employing an effective evaluation of the partial rates.

The first methods need an explicit construction of the final nuclear states
in the context of a nuclear model (shell model, RPA, Fermi gas model etc.)
and, therefore, they are, in general, more tedious. This difficulty is
avoided in the approximations of the second category (see below).

\subsubsection{State-by-state calculation of the $\mu-e$ conversion rates}
 
With this method, by constructing explicitly
the final nuclear states $\mid f \rangle$ in the context of a nuclear 
model, the branching ratio $R_{\mu e^-}$
is obtained by ``summing the partial rates'' for all possible excited states.
By using the multipole expansion of the
$\mue$ operators Eqs. (\ref{3.4.3}) and (\ref{3.4.4}),
the total rate matrix elements $M^2_{tot}$ are evaluated by
\begin{eqnarray}
 M^2_{tot} = {\sum_s}(2s+1)
f_s^2 \,\, \Big[ {\sum_{f_{exc}}} &\Big( 
{{q_{exc}} \over {{m_{\mu}}}} \Big)^2& {\sum_{l,J}}\, {\mid \langle f_{exc} 
\mid \mid {\hat T}^{(l,s)J}\mid \mid gs \rangle \mid}^2+\nonumber \\
&\Big({{ q_{gs}} \over {m_{\mu}}}\Big)^2& {\sum_{l,J}}\,
{\mid  \langle gs
\mid \mid {\hat T}^{(l,s)J}\mid \mid gs \rangle \mid}^2 \Big]
\label{5.2.1}
\end{eqnarray}
(s=0 for the vector operator and s=1 for the axial vector one). The first
term in the brackets of Eq. (\ref{5.2.1})
contains the contribution coming from all the excited states
$\mid f_{exc} \rangle$ of the final nucleus (incoherent rate) and the
second term contains the contributions
coming from the  ground state to ground state (coherent rate).

\bigskip
{\it State-by-state RPA calculations }
\bigskip

In actual shell model calculations it is quite hard to construct the final
states explicitly in medium and heavy nuclei. In such cases one can employ
the RPA approximation. In the context of the quasi-particle RPA the final
states entering the partial rate matrix elements are obtained by acting on the
vacuum $\mid 0 \rangle$ with the phonon creation operator
\beq
 Q^+(fJM) =
{\sum_{a,\tau}} \Big [ X_{a}^{(f,J,\tau)} A^+(a,JM)  -
 Y_{a}^{(f,J,\tau)} A(a,{\overline {JM} }) \Big ] 
\label{5.2.2}
\eeq
(angular momentum coupled representation) i.e. $\mid f \rangle = Q^+ 
\mid 0 \rangle$. The quantities X and Y in Eq. (\ref{5.2.2}) are the
forward and backward scattering amplitudes. The index $a$, runs over proton
($\tau=1$) or neutron ($\tau=-1$) two particle configurations coupled to J.

The transition matrix element giving the partial rate $\Gamma_{i \ra f}$
from an initial state $0^+$ to an excited state $\mid f \rangle$ takes the form
\beq
\langle f \mid \mid {\hat T}^{(l,S)J}\mid \mid 0^+ \rangle =
{\sum_{a,\tau}} \,\, W_{a}^J\,
 \Big [\,\, X_{a}^{(f,J,\tau)} U^{(\tau)}_{j_2}V^{(\tau)}_{j_1}  +
 Y_{a}^{(f,J,\tau)} V^{(\tau)}_{j_2}U^{(\tau)}_{j_1} \,\, \Big]
\label{5.2.3}
\eeq
The probability amplitudes $V$ and $U$ for the single particle
states to be occupied and unoccupied, respectively, are determined from
the BCS equations and the X and Y matrices are provided by solving 
the QRPA equations. The quantities $W_a^J \equiv W_{j_2j_1}^J$
contain the reduced matrix elements of the operators ${\hat T}$
of Eqs. (\ref{3.4.3}) and (\ref{3.4.4}) between
the single particle proton or neutron states $j_1$ and $j_2$ as
 \beq
W_{j_2j_1}^J = \langle j_2 \mid\mid {\hat T}^J\mid\mid
j_1\rangle /(2 J +1)
\label{5.2.4}
\eeq

\subsubsection{Effective computation of the individual $\mu^- \to e^-$
partial rates}

The contribution of each individual state in the total $\mu-e$ conversion
rate is effectively taking into account in the methods:
i) nuclear matter mapped into finite nuclei and
ii) sum rule approach (closure approximation).
For a short description of these methods we devote the next subsections.

\bigskip
{1. \it Nuclear matter mapped into finite nuclei via a local density 
approximation}
\bigskip

This method, uses the exact muon wave function and gives the incoherent
rate in terms of the Lindhard function~\cite{Chiang}
\begin{equation}
\bar{U}_{1,2} (q) = 2 \int \frac{d^{3} p}{(2 \pi)^{3}}
\frac{n_{1} ({\bf p}) [1 - n_{2} ({\bf q} + {\bf p})]}{q^{0} +
E_{1} ({\bf p}) - E_{2} ({\bf q} + {\bf p}) + i \epsilon}
\label{5.2.5}
\end{equation}
where $n_{1} ({\bf q}')$ and $n_{2} ({\bf q})$ the integral (0 or 1)
occupation probabilities of proton-particle proton-hole 
(or neutron-particle neutron-hole), respectively.
In the $\mu-e$ conversion this function corresponds to ph excitations of
p-p or n-n type. 

This method resembles to the relativistic Fermi gas model and
uses the local density approximation for the evaluation of
the incoherent $(\mu^-, e^-)$ conversion rate, as 
\begin{eqnarray}
\Gamma_{inc} (\mu^{-}  A \rightarrow e^{-} A^* )
& = & - 2 \int d^{3}x \vert
\Phi_{\mu} ({\bf x}) \vert^{2} \int \frac{d^{3} p_{e}}{(2 \pi)^{3}}
\Pi_{i} \frac{2 m_{i}}{2 E_{i}}\nonumber \\
&\times &
 [ \bar{\Sigma} \Sigma \vert T \vert^{2} (\mu^{-} p \rightarrow
e^{-} p) Im \bar{U}_{p,p} (p_{\mu} - p_{e})\nonumber \\
&+ &\bar{\Sigma} \Sigma \vert T \vert^{2} 
(\mu^{-} n \rightarrow e^{-} n) Im \bar{U}_{n,n} (p_{\mu} - p_{e}) ]
\label{5.2.6}
\end{eqnarray}
T represents the transition matrix, i.e. the amplitude associated
with the elementary processes $\mu^-p\ra e^-p$ and
$\mu^-n\ra e^-n$, and $m_{i}, E_{i}$ 
are the masses, energies of the particles involved in these reactions.
The two terms in the brackets of Eq. (\ref{5.2.6}) result because of the
charge conserving character of the $\mue$ operator.
 
One can separate the nuclear dependence from the dependence on the elementary 
sector in Eq. (\ref{5.2.6}) by factorizing outside the integral of 
eq. (\ref{5.2.6}) 
an average value of the quantity $\bar{\Sigma} \Sigma \vert T \vert^{2}$.
For the ratio R of the incoherent $\mue$ rate divided by the total muon 
capture rate, we get
\begin{equation}
R \, = \, \frac{\Gamma_{inc} (\mu^-,e^-)}{\Gamma (\mu^-,\nu_\mu)} =
\frac{f^{2}_{1} + f^{2}_{2}}{2} \,
\left(\frac{ \beta_0(3+\beta)}{2} \right)^2 \,G (A,Z)
\label{5.2.7}
\end{equation}
(the second step holds for the non-photonic case, involving the box diagrams
of Fig. 1(c)). All the nuclear information is contained in the function 
$G(A,Z)$ defined in Ref.~\cite{Chiang}.
Notice that $R$ differs from the branching ratio $R_{\mu e^-}$, because
R  does not include the coherent part of the $\mue$ process. 
The calculation of R in various nuclei is based upon $G(A,Z)$, since the
$\vert T \vert^{2}$ at the elementary level is in most models
independent on the nuclear parameters.
 
The branching ratio $R_{\mu e^-}$ in this method is obtained by adding in
eq. (\ref{5.2.7}) the quantity
\begin{equation}
R'=\frac{\Gamma_{coh}(\mu^- \to e^-)}{\Gamma (\mu^- \to \nu_{\mu})}
\label{5.2.8}
\end{equation}
calculated independently. To calculate $R'$ one can use the experimental
data for the nuclear form factors (electron scattering, pionic
atoms data etc.~\cite{Vries}) and the muon capture data~\cite{Suzu}.

\bigskip
{2. \it Sum rule approach }
\bigskip
 
In the closure approximation, the contribution of each final state $\mid f
\rangle$ to the total rate is also approximately taken into account.
One assumes a mean excitation energy of the nucleus 
${\bar E } = \langle E_f \rangle -E_{gs}$,
uses completeness relation for the final states $\mid f \rangle$, i.e.
$\sum _f|f \rangle \langle f| =1$, and avoids
the construction of each final state explicitly.

The method proceeds by defining the
operator taken as a tensor product from the single-particle operators ${\hat
T}$ of Eq. (\ref{3.4.3}) or (\ref{3.4.4}). For a $0^+$ initial
(ground) state the relevant tensor product is
\beq
{ \hat O }\,=\,\sum_{k,k^{\prime}} \Big[\,{\hat T}^k \times
{\hat T}^{k^{\prime}} \,\Big]^0_0 
\label{5.2.9}
\eeq
The corresponding total rate matrix elements are then written as
\beq
 M_{tot}^2=  \,
\Big( {{\mid {\bf k} \mid}\over {m_{\mu}}}\Big)^2 \, \Big[
f^2_V \langle i \mid {\hat O}_V \mid i \rangle \, + 
\, 3 f^2_A\langle i \mid {\hat O}_A \mid i \rangle \Big]
\label{5.2.10}
\eeq
where $\mid {\bf k} \mid$ is the average momentum transfer.
The operators ${\hat O}_V$ (vector) and ${\hat O}_A$
(axial vector) contain both one-body and two-body pieces.
Consequently, in the closure approximation one has to evaluate the
matrix elements of the operators ${\hat O}_V$, ${\hat O}_A$
in the ground state $\vert i \rangle$ described by a specific nuclear
model.
 
The mean excitation energy $\bar E$ of the nucleus, involved
in the matrix elements is defined as~\cite{Goul-Prim}
\beq
{\bar E} =\,\, { { {\sum_{f}} (E_f - E_{gs}) \, q_f^2\, {\mid 
\langle f \mid {\hat T}^J \mid gs \rangle \mid}^2 } \over { {\sum_f}
\, q_f^2\,
{\mid \langle f \mid {\hat T}^J \mid gs\rangle \mid}^2 }} 
\label{5.2.11}
\eeq
where the numerator represents the energy
weighted sum rule and the denominator the non-energy weighted sum rule.

Obviously, the mean
excitation energy $\bar E$ should be evaluated by constructing explicitly all
the possible excited states $\mid f \rangle$ in the context of a nuclear model.
But in the closure approximation one wishes to avoid such calculations.
Fortunately, ${\bar E}$ does not strongly depend on $(A,Z)$ and its calculation 
for some properly chosen representative nuclei 
enables the use of closure for every nucleus
throughout the periodic table.
In Ref.~\cite{KVCF}, the QRPA method has
been used for the determination of the mean excitation energy 
for some nuclear isotopes including $^{48}Ti$ and $^{208}Pb$
which are of current experimental interest. 
 
\section{Results and Discussion}

\subsection{Nuclear structure dependence of the $\mu^- \to e^-$
conversion rates}
 
The nuclear physics calculations done for the $(\mu^-,e^-)$ conversion
mainly refer to the most interesting quantities of the process, i.e.
the branching ratio $R_{\mu e^-}$ and the ratio $\eta$ of Eq. (\ref{4.1}).
Several nuclear methods for the most important 
mechanisms leading to this process have been employed.
A general feature of the results obtained for various types of diagrams
is the fact that qualitatively they are very similar
(see as an example Fig. 3 below). The quantitative differences are mainly
related to the presence or not of neutron contributions
(they exist only for non-photonic diagrams) and the different values
of the parameters $\beta, \beta '$ and $\beta ''$ of Eqs.
(\ref{3.1.5}), (\ref{3.1.6}) and (\ref{3.1.7})
which are the only elementary
physics parameters entering the nuclear transition matrix elements.

For the coherent rate and the total rate obtained using a sum rule
approach, only the ground state of the nucleus need be constructed. 
The first such extensive calculations for closed (sub)shell nuclei
were done with a determinantal ground state wave function (Slater 
determinant). These calculations provided a good 
description of the coherent rate~\cite{KV90}. The sum-rule results, 
however, have shown that the total rates are sensitive to
the mean excitation energy required in the closure approximation.
Furthermore, the local density approximation~\cite{Chiang} gives 
reliable results for the
coherent rate mainly due to the exact computation of the muon-nucleus
overlap integral. The main advantage of this method in calculating
the incoherent rate (by integrating over a continuum of excited states
of a local Fermi sea), is the fact that it takes into account the
contributions coming from the continuum, which are not explicitly 
included in RPA and shell model results.

Below we discuss the results obtained in the framework of the
quasi-particle RPA and discuss their comparison with those of shell
model and local density approximation. The coherent and incoherent
rate are discussed separately.

\subsubsection{ The coherent process}

As we have mentioned in sect. 5, for the coherent rate
the proton, $F_Z(q^2)$, and neutron, $F_N(q^2)$, nuclear form factors
are required. The results obtained by using quasiparticle RPA and shell 
model are shown in Fig. 3. They refer to the following two cases: 
(i) By neglecting the muon binding energy $\epsilon_b$ in Eq. (\ref{3.4.5}) 
the elastic momentum transfer is the same for all nuclei,
i.e. $q
\approx m_{\mu} \approx 0.535 fm^{-1}$ (results denoted QRPA(i)).
(ii) By taking into account $\epsilon_b$ in Eq. (\ref{3.4.5}), the 
elastic momentum transfer is $q \approx m_{\mu} - \epsilon _b$ and
varies from $q \approx 0.529 fm^{-1}$, for $^{48}Ti$ where $\epsilon_b
\approx 1.3 MeV$, to  $q \approx 0.482 fm^{-1}$, for $^{208}Pb$ where
$\epsilon_b \approx 10.5 MeV$ (results indicated as QRPA(ii)).
From the variation of the coherent nuclear matrix elements $M^2_{coh}$ with
respect to A and Z illustrated in Fig. 3(a) (photonic mechanism), and
Fig. 3(b) (non-photonic mechanism), we see that
by taking into consideration the muon binding energy $\epsilon_b$,
QRPA(ii), all matrix elements increase continuously up to the $Pb$ region
where they become about a factor of two larger than the approximate
results of case (i). This implies that the coherent rate becomes larger for
heavy nuclei, $Pb$ region, which makes such nuclei attractive from an 
experimental point of view~\cite{Honec,Ahmad} provided, of course,
that they satisfy other additional criteria, as
the minimization of the reaction background etc.~\cite{Depom,Walt}. The
$\mue$ conversion electrons of a given target are expected to show a
pronounced peak around $E_e = m_{\mu}- \epsilon _b$ (for lead 
region $E_e \approx 95 MeV).$  

The above results have been obtained using
the factorization approximation Eq. (\ref{5.1.6}). 
This approximation only slightly 
affects the ratio $\eta$ of the coherent rate to the total rate.
However, the branching ratio $R_{\mu e}$ for heavy nuclei is appreciably
affected by using exact muon wave function. This gives 
results which for heavy nuclei are $30-40 \%$ larger than the
approximate ones. 

We should also note that, it is important to estimate
the effect of the ground state RPA
correlations on the $\mue$ matrix elements.
This can be done by using a correlated quasi-particle RPA
vacuum instead of the uncorrelated one. In Ref.~\cite{KVCF}
such calculations have been performed using
the correlated vacuum defined by~\cite{Sande,Row}
\beq
 {\mid {\tilde 0 } \rangle}\,=\,\, N_0\,e^{{\hat S}^+}{\mid 0\rangle} 
\label{6.1.1}
\eeq
where the operator ${\hat S}^+$ contains the correlation matrix
and $N_0$ is an appropriate normalization factor.
In this case, the coherent rate matrix elements take the form
\beq
\langle {\tilde 0} \mid {\hat T}\mid {\tilde 0} \rangle \,
= \,{ N^2_0}\, \langle 0 \mid {\hat T} \mid 0 \rangle
\label{6.1.2}
\eeq
which means that the correlated matrix elements are 
obtained by rescaling the
uncorrelated ones (a similar expression holds for the total
rate matrix elements).
It was found that the matrix elements for $^{48}Ti$
obtained this way~\cite{KVCF} are reduced by $\approx 30 \%$
and that the ground state
correlations tend to decrease the strengths of all $\mue$ conversion channels.

\subsubsection{ Incoherent process}

The results for the incoherent rate 
obtained with ``state-by-state'' QRPA calculations of all the excited
states for a series of nuclei throughout the periodic table are shown
in Table 2 for the photonic diagrams, and Fig. 4
for the non-photonic ones.
Positive and negative parity states up to $6^-$, $6^+$ are included.
For the photonic mechanism only the vector component, $S_V$ gives non-zero
contribution ($M^2_{inc} = S_V $).
For the non-photonic mechanism we have non zero
contributions from both the vector and axial vector components, $S_V$ and
$S_A$, ($M^2_{inc} = S_V + 3S_A$). 

It can be seen that, the main contribution to the incoherent
rate comes from the $1^-$ multipolarity. For every multipole
only the low-lying excited states give significant contributions.
High-lying excited states contribute negligibly.
Furthermore, the incoherent matrix elements do not show clear A and Z
dependence. Their magnitude depends on the spectrum of the individual
nuclear isotope.
  
In obtaining the results of Table 2 and Fig. 4 for $1^-$ states, 
the spurious center of mass contaminations, which arise from the 
use of empirical single particle energies and the truncation of
the model space, have been eliminated by explicitly calculating the purely
spurious state $|S>$ and removing its admixtures from the incoherent and
total rates. The calculation of the overlaps $\langle 1^-,m|S\rangle$ (where
$m$ counts the $1^-$ contaminated excited states) have shown that mostly the
spuriousity 
lies in the lowest $1^-$ state being $\sim 80\%$ for $^{48}Ti$ and
$^{208}Pb$ (for other nuclei see Ref.~\cite{SFK97}). 
We showed that in a good
approximation we can neglect the first $1^-$ state, but then we have
to renormalize the strengths of the residual interaction (Bonn
potential).  
From the renormalization we found that the incoherent
$\mu^- \to e^-$ conversion rate is not sensitive to the strength
parameters of the two-body interaction, which increases the
reliability of the RPA results.
For all nuclei studied, the spurious center of mass
contribution is less than 30\% of the incoherent matrix elements,
i.e., $\sim 2\%$ of the total $\mue$ conversion rate.
Furthermore, even after removal of the spurious components the $1^-$
multipolarity appears to be the most important incoherent channel.

The explicit construction of all final nuclear states enables the
calculation of the mean excitation energy ${\bar E}$ needed in the closure
approximation. Though ${\bar E}$ is defined in analogy with the
ordinary muon capture reaction by Eq. (\ref{5.2.11}), the values of the 
``mean excitation energies'' obtained in the $\mue$ are completely different
than those evaluated in $(\mu^-,\nu_{\mu})$~\cite{KVCF}
because in the last process the coherent
channel doesn't exist. For the $\mue$ process the coherent channel 
appears only in the denominator of Eq. (\ref{5.2.11}) and,
since this dominates the total $\mu-e$ conversion rate, the resulting mean
excitation energy $\bar E$ in this process is  much smaller
than that for the $(\mu^-,\nu_{\mu})$ reaction (see Table 3). 

\subsection{Comparison of coherent and incoherent processes}

As we have seen in sect. 4, a useful quantity for the $\mue$ conversion is
the fraction $\eta$ of the coherent matrix elements $M^2_{coh}$ divided by 
the total one $M^2_{tot}$.
In earlier calculations $\eta$ was estimated~\cite{WeiFei} to be a
decreasing function of A with a value of $\eta \approx$ 83 \% in Cu region. By
using, however, local density approximation~\cite{Chiang} and 
QRPA~\cite{KFSV97} we find
that the coherent channel dominates throughout the periodic table.

The values of $\eta$ obtained with QRPA are a bit larger than
the more accurate results obtained with a local density 
approximation~\cite{Chiang} (LDA) in the medium and heavy nuclei region.
The reason is the following: The LDA, for the part of
non-coherent contributions, uses neither closure nor explicit
summation over all final states. It makes a summation over
the continuum of excited states (even though they contribute negligibly
they are infinite in number) in a local Fermi sea.
The incoherent rate obtained this way, is tied to the number of states
one can excite; the larger the better; 
the muon mass provides the energy for the excitation of
such states. Consequently, LDA takes into account the contribution
of the continuum spectrum and calculates it more accurately for heavier
nuclei. Such contributions are absent from the RPA results for every
nucleus.

\subsection{Limits of the elementary sector part of $R_{\mu e^-}.$
Calculation of $\gamma(A,Z)$}
 
The task of studying the nuclear physics aspects of the exotic
$\mue$ conversion process is to evaluate the function $\gamma (A,Z)$ 
of Eqs. (\ref{4.3}), (\ref{4.1.1}), (\ref{4.1.9}) and (\ref{4.1.11}). 
In general, the nuclear matrix elements entering $\gamma(A,Z)$ 
depend on the final nuclear state populated during process (3).
The values for $\gamma(A,Z)$ calculated from the coherent QRPA matrix elements 
are shown in Table 4 and compared with the results of 
Refs.~\cite{KV90,Chiang}.
The variation of $\gamma (A,Z)$ through the periodic
table~\cite{KV90} exhibits a strong dependence on the neutron
excess $(N-Z)$ which mainly reflects the dependence on $(A,Z)$
of the total muon capture rate.
We should note that, the muon binding energy $\epsilon_b$ of Eq. 
(\ref{3.4.5}), which affects significantly the nuclear form factors for 
heavy nuclei like $^{208}Pb$, affects also the factor $E_e p_e/m^2_\mu$
in the definition of $\gamma$ Eq. (\ref{4.3}). This factor takes into
account the phase space in the transition matrix elements.
By ignoring $\epsilon_b$, this factor becomes unity. 
For $^{48}Ti$ this factor, by including $\epsilon_b$, is equal to 0.97 
and its neglect is a good approximation. 
For $^{208}Pb$, however, this factor is equal to
0.81 which means that, in heavy nuclei the dependence on $\epsilon_b$ 
of the phase space cannot be ignored.

By putting the experimental limits~\cite{Walt,Honec} of the branching ratio 
$R_{\mu e^-}$ on Eq. (\ref{4.2}) 
we determine upper bounds on the parameter $\rho$ for photonic, Z-exchange and 
W-box diagrams both in conventional extensions of the standard model as well 
as SUSY theories (for the currently interesting nuclei $^{48}Ti$
and $^{208}Pb$ see Table 4).
We note that the values of $\rho$ for $^{48}Ti$
are slightly improved over those of Table 2 of Ref.~\cite{KVCF}, but those of 
$^{208}Pb$ are appreciably smaller than those of sect. 5.1 of Ref.~\cite{KV90}.
This big difference is due to the fact that, in the present work we have used 
the experimental limit of Ref.~\cite{Honec}. This experiment at PSI improved 
on the previous limit~\cite{Ahmad} of $^{208}Pb$, which was used 
in Ref.~\cite{KV90}, by an order of magnitude. For both nuclei we found
that, the 
limits obtained from various nuclear models do not significantly differ from 
each other. We should stress, however, that, the different limits for
$^{208}Pb$ in the 
shell model results of Table 4, are due to the neglect of $\epsilon_b$ in Eq. 
(\ref{3.4.5}) when calculating the nuclear form factors as mentioned above; 
its consideration gives 
similar results to those of LDA and QRPA for both mechanisms. This implies 
that all nuclear models studied here give about same values for $\rho$.

One can use the limits of $\rho$ to parametrize the muon number violating 
quantities entering Eqs. (\ref{4.1.2}), (\ref{4.1.4}), (\ref{4.1.10}) and
(\ref{4.1.12}) and also to compare them
directly to the value given from gauge models. As an example, we quote the value
of $\rho = 8.2 \times 10^{-18}$ obtained in the supersymmetric model of 
Ref.~\cite{KV90} discussed above. This prediction of $\rho$ is considerably 
smaller compared to the values listed in Table 4, which were extracted from 
experiment. 

\section{ Summary and Conclusions}

The main conclusions stemming out of the results discussed above
can be summarized as follows.
The coherent mode dominates throughout the periodic table
indicated from the fact that $\eta \ge 90 \%$.

The quasi-particle RPA and Local Density Approximation
results for the $\mue$ rate do not show maximum around 
$A \approx 60$ as had previously estimated. This rate keeps increasing
up to heaviest nuclei, i.e. region of $^{208} Pb$, presently
used at PSI as target.
In evaluating the nuclear matrix elements the muon binding energy
should not be ignored especially for heavy nuclei.

The great part of the incoherent rate comes from the low-lying 
excitations. The $1^-$ state gives the maximum incoherent contribution even 
after removing the spurious admixtures.

In contrast to the ordinary muon capture, the mean excitation 
energy, $\bar E$, of the nucleus in $\mu^- \, \ra \, e^-$ is very small, 
${\bar E} \approx 1 MeV$.

The dependence of $R_{\mu e^-}$ on nuclear physics in coherent production,
expressed by the function $\gamma(A,Z)$ shows a large
variation. Thus, for
example, in the case of photonic mechanism, $\gamma$ lies in the region
$1.6 \le \gamma_{ph} \le 26.0$ from lighter to heavier nuclei.

All nuclear models used give about the same values for $\gamma$ and
quantity $\rho$ which contains the elementary sector parameters
of the branching ratio $R_{\mu e^-}$. The quantity $\rho$ is useful
to fix the lepton flavour violating parameters and test the various 
gauge models.

The observation of any muon number (and in general any lepton and/or
flavour number) non-conserving process, will reveal ``new physics'' beyond 
the standard model. The predicted branching ratios, however, are much 
smaller than experimental limits.
The most optimistic results are obtained in the context of supersymmetry
$1.2 \times 10^{-18} \le R_{\mu e^-} \le 2.4 \times 10^{-16}$

Although the predicted branching ratios for $\mu^- \ra e^-$ process
is much smaller than present-day experimental sensitivity,
this should not discourage the relevant experiments, since it is a common
belief that we do not as yet have a complete theory to adequately describe
such exotic processes.

\bigskip
\noindent
T.S.K. is grateful to the Organizing Committee of the Conference for 
financial support and the hospitality at Dubna.

\bigskip

\newpage
\noindent
{\bf Table 1.} Best upper limits for the branching ratios of some muon
number violating processes: Elementary particle decay modes and exotic
neutrinoless muon decays in a nucleus.

\begin{center}
\begin{tabular}{|lclc|}
\hline
\hline
& & & \\
Process & \multicolumn{1}{r}{ Best upper limit} & 
   \multicolumn{1}{l}{ Reaction } &
   \multicolumn{1}{l|}{ Ref. } \\
\hline
& & & \\
\multicolumn{4}{|c|}{ \bf Elementary muon number violating
processes } \\
\hline
& & & \\
$ \mu \to e\gamma$ & $R_{e\gamma} < 4.9\times 10^{-11}$ & & [36] \\
$ \mu \to e e^+ e^- \quad (\mu \to 3e)$ & 
$R_{3e} < 1.0\times10^{-12}$ & 
& [37] \\
$\tau \to \mu\gamma$ &
$R_{\tau} < 4.2\times10^{-6}$ & & [38] \\
$ \tau \to \mu e^+ e^-, $ & & & \\
$ \tau \to e \mu^+ \mu^- $ & & & \\
$ \tau \to \mu \mu^+\mu^-$ & & & \\
$ (\mu^+ e^-) \leftrightarrow (\mu^- e^+)$ & 
$R_{M \bar M} < 6.5\times10^{-7}$ & & [39] \\
\hline
& & & \\
\multicolumn{4}{|c|}{ \bf Neutrinoless rare muon decays in 
a Nucleus } \\
\hline
& & & \\
$\mu^- (A,Z) \to e^- (A,Z) ^*$ & $R_{\mu e^-} < 7.0\times10^{-13}$&
${\mu^-Ti \to e^-Ti}$ & [5] \\
&$ R_{\mu e^-} < 4.9\times10^{-11}$ &
${\mu^-Pb \to e^-Pb}$ & [6] \\
$\mu^- (A,Z) \to e^+ (A,Z-2) $ & $R_{\mu e^+} < 5.5\times10^{-12}$ &
$\mu^- Ti \to e^+Ca$ & [22] \\
\hline
\hline
\end{tabular}
\end{center}

\noindent
{\bf TABLE 2.} Incoherent $\mu -e$ conversion matrix elements ($M^2_{inc}$) 
for the photonic mechanism. Only the vector component $(S_V)$ of the 
$\mu - e$ conversion operator (see Eq. (\ref{3.4.1})) contributes. 

\vskip0.3cm 

\begin{center}
\begin{tabular}{|c|llllll|}
\hline
\hline
        &         &        &         &       &          &         \\
$J^{\pi}$&$^{48}_{22}Ti$&$^{60}_{28}Ni$&$^{72}_{32}Ge$ 
&$^{112}_{48}Cd$&$^{162}_{70}Yb$&$^{208}_{82}Pb$\\
\hline
        &         &        &         &       &          &         \\
$ 0^+ $ & 1.946   & 1.160  &  2.552  &  2.088  &  4.305 &  2.512  \\
$ 2^+ $ & 0.242   & 0.738  &  1.396  &  2.669  &  6.384 &  2.342  \\
$ 4^+ $ & 0.004   & 0.005  &  0.015  &  0.021  &  0.063 &  0.056  \\
$ 6^+ $ &6 $10^{-6}$&6 $10^{-6}$&1 $10^{-5}$&6 $10^{-5}$  & 2 $10^{-4}$ 
 & 3 $10^{-4}$  \\
        &         &         &        &         &        &         \\
$ 1^- $ & 3.711   & 4.215   & 5.066  &  5.282  &  4.824 &  4.533  \\
$ 3^- $ & 0.037   & 0.081   & 0.152  &  0.249  &  0.542 &  0.476  \\
$ 5^- $ &2 $10^{-4}$&2 $10^{-4}$&5 $10^{-4}$& 0.001&0.005 &  0.005  \\
\hline
        &         &         &        &         &        &         \\
 $M^2_{inc}$ & 5.940  &  6.199 &  9.181 & 10.309 & 16.123 &  9.924 \\
\hline
\hline
\end{tabular}
\end{center}

\vspace{5mm}
{\bf Table 3.} Mean excitation energies for some representative nuclei
in the reaction: $\mu^-_b + (A,Z) \to e^- + (A,Z)^*$. In contrast to the
ordinary muon capture reaction, where ${\bar E} \approx 20 \, MeV$, in 
the $\mu - e$ conversion ${\bar E}$ is very smaller.
\begin{center}
\begin{tabular}{|l|cccccc|}
\hline
\hline
        &         &        &         &       &          &         \\
Nucleus&$^{48}Ti$&$^{60}Ni$&$^{72}Ge$ &$^{112}Cd$&$^{162}Yb$&$^{208}Pb$\\
\hline
          &        &        &        &        &        &         \\
${\bar E} \, (MeV)$&  0.84  &  0.44  &  0.50  &  0.40  &  0.44  &  0.28   \\
\hline
\hline
\end{tabular}
\end{center}

\vspace{5mm}
{\bf Table 4.} The new limits on the elementary sector part of the exotic 
$\mu-e$ conversion branching ratio extracted by using Eq. (\ref{4.2}) and
the recent experimental data for the nuclear targets $^{208}Pb, \,
\,\,^{48}Ti$,~\cite{Walt,Honec}.
The nuclear part of the branching ratio, described by the 
function $\gamma (A,Z)$ (see sect. 4), is also shown.

\vskip0.8cm
\begin{center}
\begin{tabular}{|llrcrc|}
\hline
\hline
& & & & & \\
& & \multicolumn{2}{c}{\large \bf $^{48}Ti$}&
    \multicolumn{2}{c|}{\large \bf $^{208}Pb$} \\
\hline
& & & & & \\
Method & Mechanism & $\gamma(A,Z)$ & $\rho$ $(\times 10^{-13}$ &
$\gamma(A,Z)$ & $\rho$ $\times 10^{-12}$ \\
\hline
& & & & & \\
    & Photon exch.    &  9.42 & $\le$ 4.6  
		      & 17.33 & $\le$ 2.7  \\
QRPA&W-boson exch.    & 25.31 & $\le$ 1.7  
		      & 49.22 & $\le$ 0.9  \\
    & SUSY s-lept.    & 25.61 & $\le$ 1.7  
                      & 49.50 & $\le$ 0.9  \\
    & SUSY Z-exch.    &110.57 & $\le$ 0.4  
                      &236.19 & $\le$ 0.2  \\
\hline
& & & & & \\
LDA & Photonic        &  9.99 & $\le$ 4.3  
         	      & 17.84 & $\le$ 2.6  \\
    &W-boson exch.    & 26.60 & $\le$ 1.6  
		      & 55.53 & $\le$ 0.8  \\
\hline
& & & & & \\
SM & Photonic         &  9.74 & $\le$ 4.4  
                      & 10.42 & $\le$ 4.4  \\
    &W-boson exch.    & 26.50 & $\le$ 1.6  
                      & 27.43 & $\le$ 1.7  \\
\hline
\hline
\end{tabular}
\vspace{7mm}
\end{center}

\newpage
\centerline{\large \bf FIGURE  CAPTIONS}

\vspace{5mm}

\noindent
{\bf FIGURE 1.} Typical diagrams entering the neutrinoless $\mue$ conversion:
photonic 1(a), Z-exchange 1(a),(b) and W-boson exchange 1(c), for the specific
mechanism involving intermediate neutrinos.

\vspace{5mm}

\noindent
{\bf FIGURE 2.} SUSY diagrams leading to the $\mue$ conversion: photonic 2(a), 
Z-exchange 2(a),(b) and box diagrams 2(c), in a supersymmetric model with 
charged s-lepton and neutralino mixing are shown. Note that, the Z-exchange in 
2(b) as well as in Fig. 1(b), comes out of electrically neutral particles 
(photon exchange does not occur in these diagrams). These Z-exchange diagrams 
may be important (those of Figs. 1(a) and 2(a) are suppressed by 
$m^2_{\mu}/m^2_Z$).

\vspace{5mm}

\noindent
{\bf FIGURE 3.} Variation of the coherent $\mue$ conversion matrix elements 
$M^2_{coh}$ for specific mass A and charge Z (see text) for the photonic
mechanism (a) and the non-photonic mechanism (b). 
In QRPA(i) the muon binding energy $\epsilon_b$ was neglected, but it was
included in QRPA(ii). We see that $\epsilon_b$ strongly affects the matrix
elements for heavy nuclei. For comparison the results of Ref.~\cite{KV90}
(shell model results) are also shown. For photonic and non-photonic diagrams
the coherent rate increases up to $Pb$ region where it starts to decrease.

\vspace{5mm}

\noindent
{\bf FIGURE 4.} Incoherent non-photonic $(\mu^-,e^-)$ conversion
rate for the most interesting nuclei from an experimental point of
view; $^{208}Pb$ is currently used as a target in the SINDRUM II
experiment at PSI and $^{48}Ti$ has provided the most stringent limit
on the flavour violation. The bars show the partial contribution 
of each multipolarity of the $\mu \to e$ conversion operator (black bars
for the vector part ($S_V$) and empty bars for the axial vector part
($S_A$)). The $1^-$ contribution shown is obtained after removing the
spurious contaminations but still it is the most important one for
this process.


\begin{thebibliography}{99}
\bibitem{Engf}R. Engfer and H.K. Walter, Ann. Rev. Nucl. Part. Sci. {\bf 36}
(1986) 327.
\bibitem{JDV}J.D. Vergados, Phys. Reports {\bf 133} (1986) 1.
\bibitem{KLV94}T.S. Kosmas, G.K. Leontaris and J.D. Vergados,
 Prog. Part. Nucl. Phys. {\bf 33} (1994) 397.
\bibitem{Depom}P. Depommier and C. Leroy, Rep. Prog. Phys. {\bf 58} (1995) 61.
\bibitem{Walt}H.K. Walter, in these proceedings, and Private communication.
\bibitem{Honec}W. Honecker {\it et al.}, (SINDRUM II Collaboration),
Phys. Rev. {\bf Lett. 76} (1996) 200.

\bibitem{WeiFei}S. Weinberg and G. Feinberg, Phys. Rev. {\bf Lett. 3}
(1959) 111; {\it ibid} E 244.
\bibitem{MarSan}W.J. Marciano and A.I. Sanda, Phys. Rev. {\bf Lett. 38}
 (1977) 1512.
\bibitem{Li}T.P. Cheng and L.F. Li, Phys. Rev {\bf D 20} (1979) 1608.
\bibitem{Altar}G. Altareli {\it et al.}, Nucl. Phys. {\bf B 125} (1977) 285. 
\bibitem{Shank}O. Shanker, Phys. Rev {\bf D 20} (1979) 1608.
\bibitem{Berna}J. Bernabeu, E. Nardi and D. Tommasini, Nucl. Phys. {\bf B 409}
(1993) 69. 
\bibitem{KLV89}T.S. Kosmas, G.K. Leontaris and J.D. Vergados,
Phys. Lett. {\bf B 219} (1989) 457.
\bibitem{KV96}T.S. Kosmas and J.D. Vergados, Phys. Reports {\bf 264}
(1996) 251.

\bibitem{muplus}J.D. Vergados and M. Ericson, Nucl. Phys. {\bf B 195} (1982)
262; S. Pittel and J.D. Vergados, Phys. Rev. {\bf C 24} (1981) 2343;
G.K. Leontaris and J.D. Vergados, Nucl. Phys. {\bf B 224} (1983) 137;
J.D. Vergados, Phys. Rev. {\bf D 23} (1981) 703.

\bibitem{Stein}J. Steinberger and H. Wolfe, Phys. Rev {\bf 110} (1955) 1490.
\bibitem{Conver}M. Conversi {\it et al.}, Phys. Rev {\bf 122} (1961) 687.
\bibitem{Brym72}D.A. Bryman {\it et al.}, Phys. Rev {\bf Lett. 28} (1972) 1469.
\bibitem{Bader82}A. Badertscher {\it et al.}, Nucl. Phys. {\bf A 377}
(1982) 406.
\bibitem{Brym85}D.A. Bryman {\it et al.}, Phys. Rev. {\bf Lett. 55} (1985) 465.
\bibitem{Ahmad}S. Ahmad {\it et. al.}, (TRIUMF Collaboration), Phys. Rev. 
{\bf Lett. 59} (1987) 970; Phys. Rev. {\bf D 38} (1988) 2102.
\bibitem{Bad91}A. Badertscher {\it et al.}, J. of Phys. {\bf G 17} (1991) S47.
\bibitem{Dohm}C. Dohmen {\it et al.}, (SINDRUM II Collaboration),
Phys. Lett. {\bf B 317} (1993) 631.
\bibitem{Scha}A. van der Schaaf, Nucl. Phys. {\bf A 546} (1992) 421c; 
Prog. Part. Nucl. Phys.  {\bf 31} (1993) 1. 

\bibitem{KV88}T.S. Kosmas and J.D. Vergados, Phys. Lett. {\bf B 215}
(1988) 460.
\bibitem{KV89} T.S. Kosmas and J.D. Vergados, Phys. Lett. {\bf B 217}
(1989) 19.
\bibitem{KV90}T.S. Kosmas and J.D. Vergados, Nucl. Phys. {\bf A 510}
 (1990) 641.
\bibitem{Chiang}H.C. Chiang, E. Oset, T.S. Kosmas, A. Faessler and
J.D. Vergados, Nucl. Phys. {\bf A 559} (1993) 526.
\bibitem{KVCF}T.S. Kosmas, J.D. Vergados, O. Civitarese and A. Faessler,
Nucl. Phys. {\bf A 570} (1994) 637.
\bibitem{KFV97}T.S. Kosmas, A. Faessler and J.D. Vergados, 
J. of Phys. {\bf G 23} (1997) 693.
\bibitem{KFSV97}T.S. Kosmas, A. Faessler, F. \v Simkovic and J.D. Vergados, 
Phys. Rev. {\bf C 56} (1997) 526.
\bibitem{SFK97}J. Schwieger, A. Faessler, and T.S. Kosmas,
Phys. Rev. {\bf C 57}, in press.
\bibitem{Scheck}F. Scheck, Phys. Reports {\bf 44} (1978) 187. 

\bibitem{Goul-Prim}B. Goulard and H. Primakoff, Phys. Rev. {\bf C 10}
(1974) 2034; 

D. Duplain, B. Goulard, and J. Joseph, Phys. Rev. {\bf C 12} (1975) 28;
\bibitem{Suzu}T. Suzuki, D. Mearsday and J. Roalsvig, Phys. Rev. {\bf C 35} 
(1987) 236.
\bibitem{Bolto}R.D. Bolton {\it et al.}, Phys. Rev. {\bf D 38} (1988) 2077.
\bibitem{Bellga}U. Bellgardt {\it et al.}, SINDRUM Collaboration, Nucl. Phys. 
{\bf B 299} (1988) 1.
\bibitem{Bean}A. Bean {\it et al.}, Phys. Rev. {\bf Lett. 70} (1993) 138.
\bibitem{Matt}B.E. Matthias {\it et al.}, Phys. Rev. {\bf Lett. 66} (1991)
2716.
\bibitem{Koval}S. Kovalenko, in these proceedings.
\bibitem{KVprd}T.S. Kosmas and J.D. Vergados, Phys. Rev. {\bf D 55}
(1997) 1752.
\bibitem{LagKos}T.S. Kosmas {\it et al.}, in preparation.
\bibitem{KV92} T.S. Kosmas and J.D. Vergados, Nucl. Phys.
 {\bf A 523} (1992) 72.
\bibitem{Heisen}J. Heisenberg, R. Hofstadter, J.S. McCarthy and I. Sick,
Phys. Rev. {\bf Lett. 23} (1969) 1402;

T.W. Donnelly and J.D. Walecka, Ann. Rev. Nucl. Sci. {\bf 25} (1975) 329;

B. Frois and C.N. Papanicolas, Ann. Rev. Nucl. Part. Sci. {\bf 37} (1987) 133.
\bibitem{Vries} H. de Vries, C.W. de Jager and C. de Vries, Atomic Data
 and Nuclear Data Tables {\bf 36} (1987) 495.
\bibitem{Sande}E.A. Sanderson, Phys. Lett. {\bf 19} (1965) 141;
 
J. Da Providencia, Phys. Lett. {\bf 21} (1966) 668.
\bibitem{Row} D.J. Rowe, Nuclear collective motion, (Methuen and CO. LTD., 
London, 1970).
\end{thebibliography}
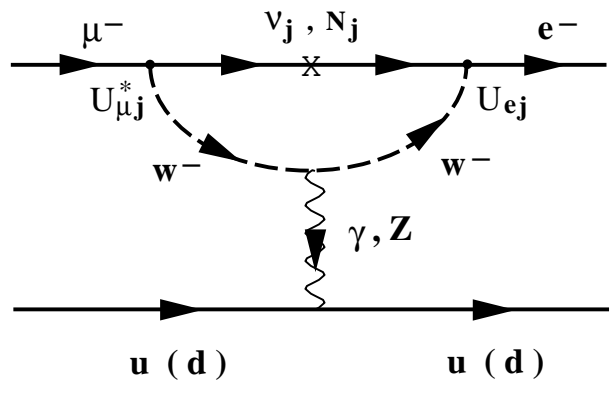
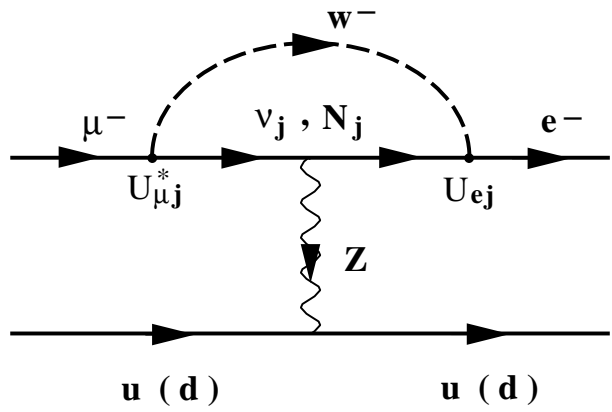
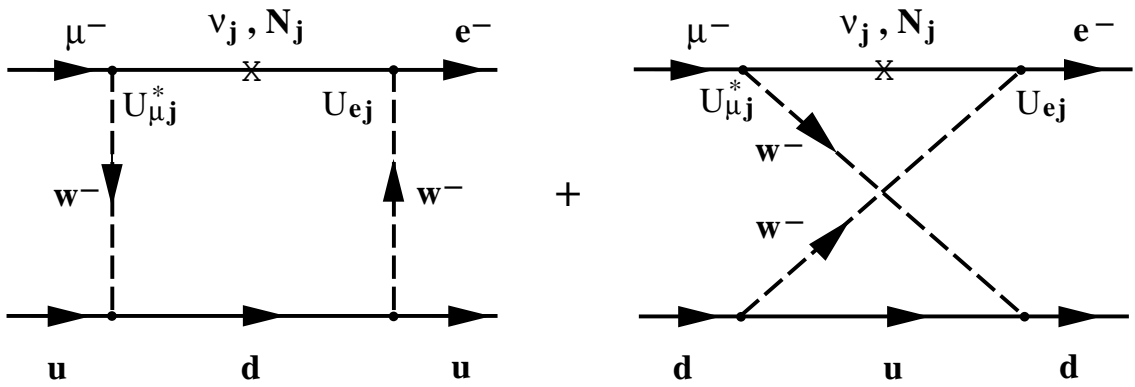
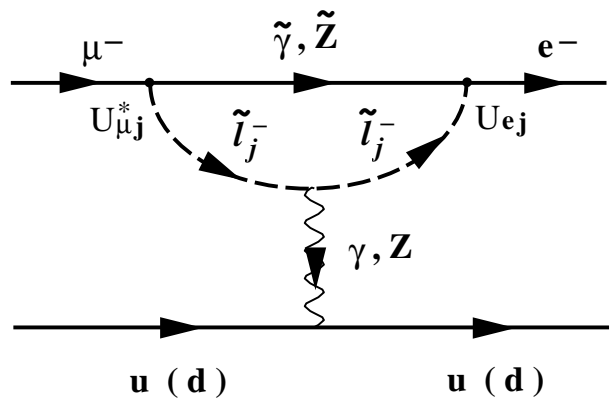
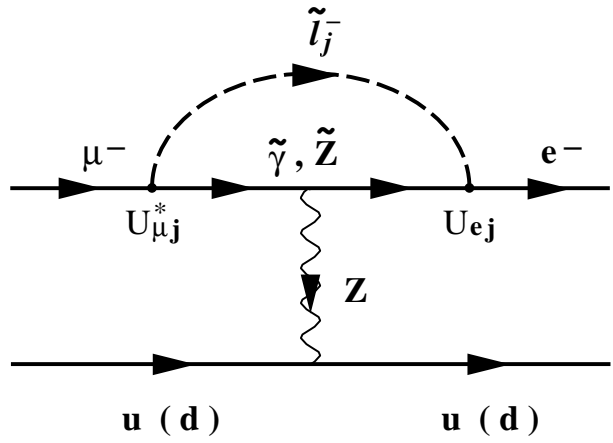
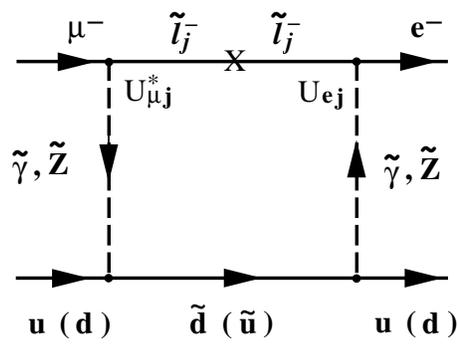
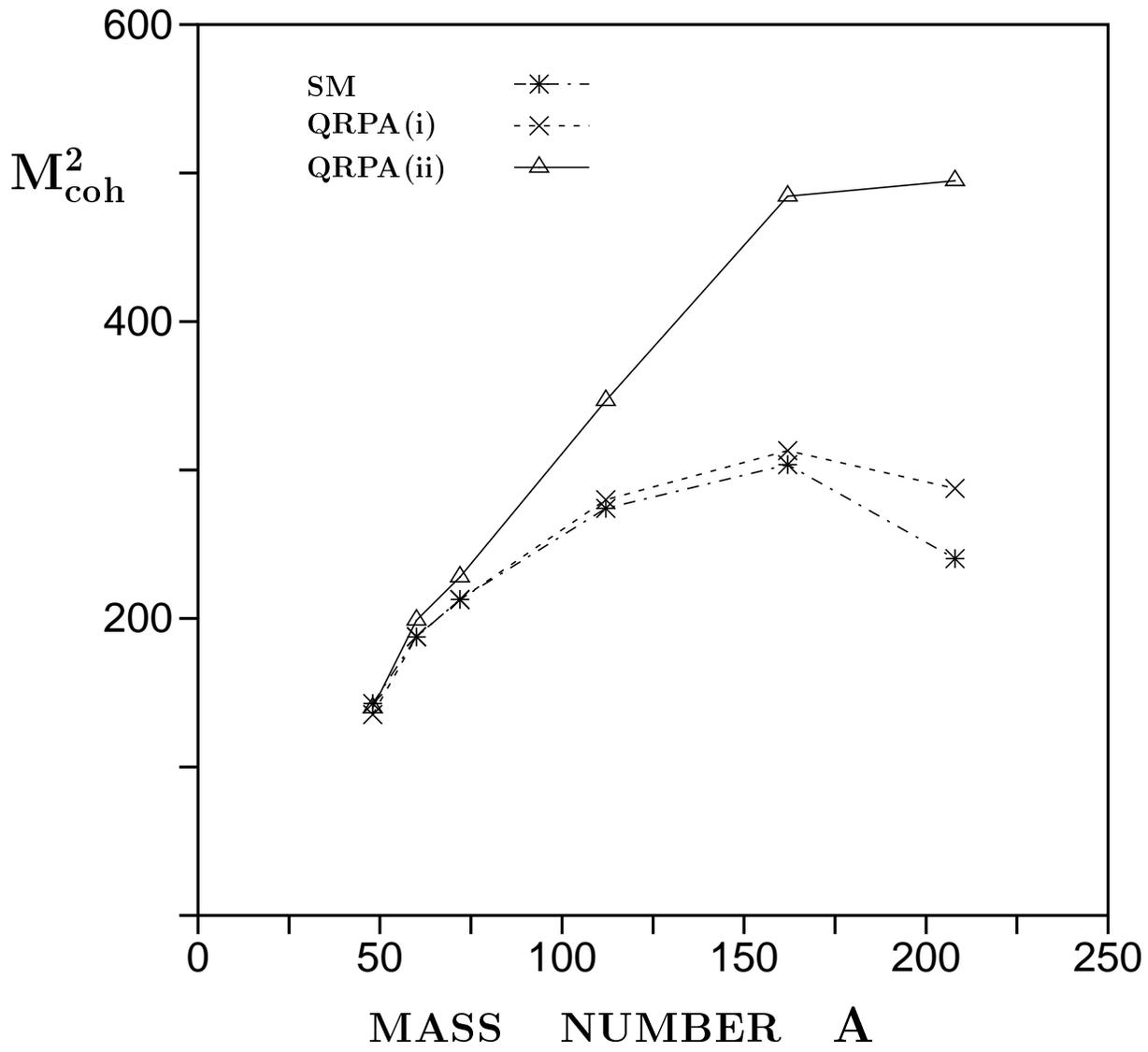
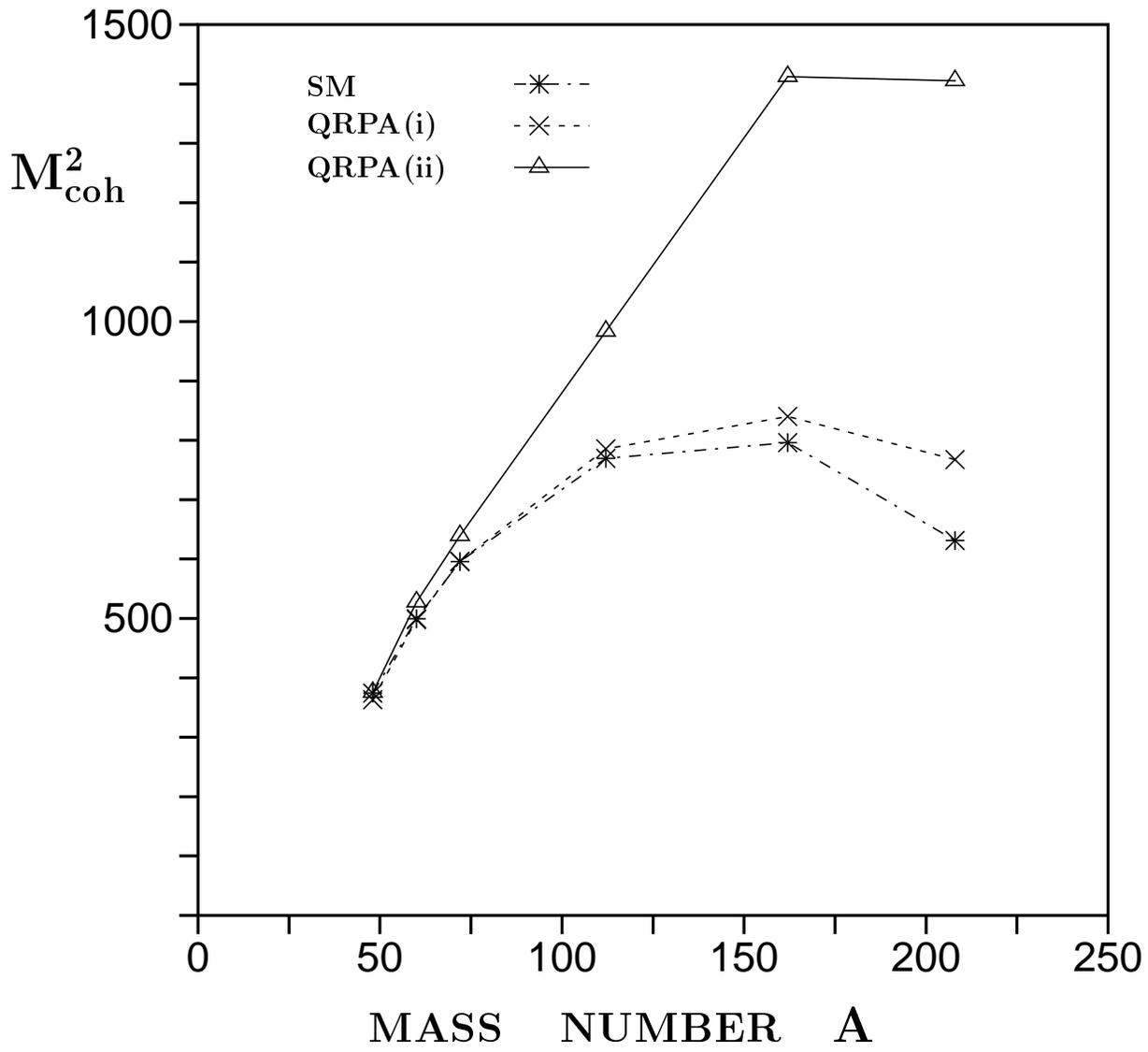
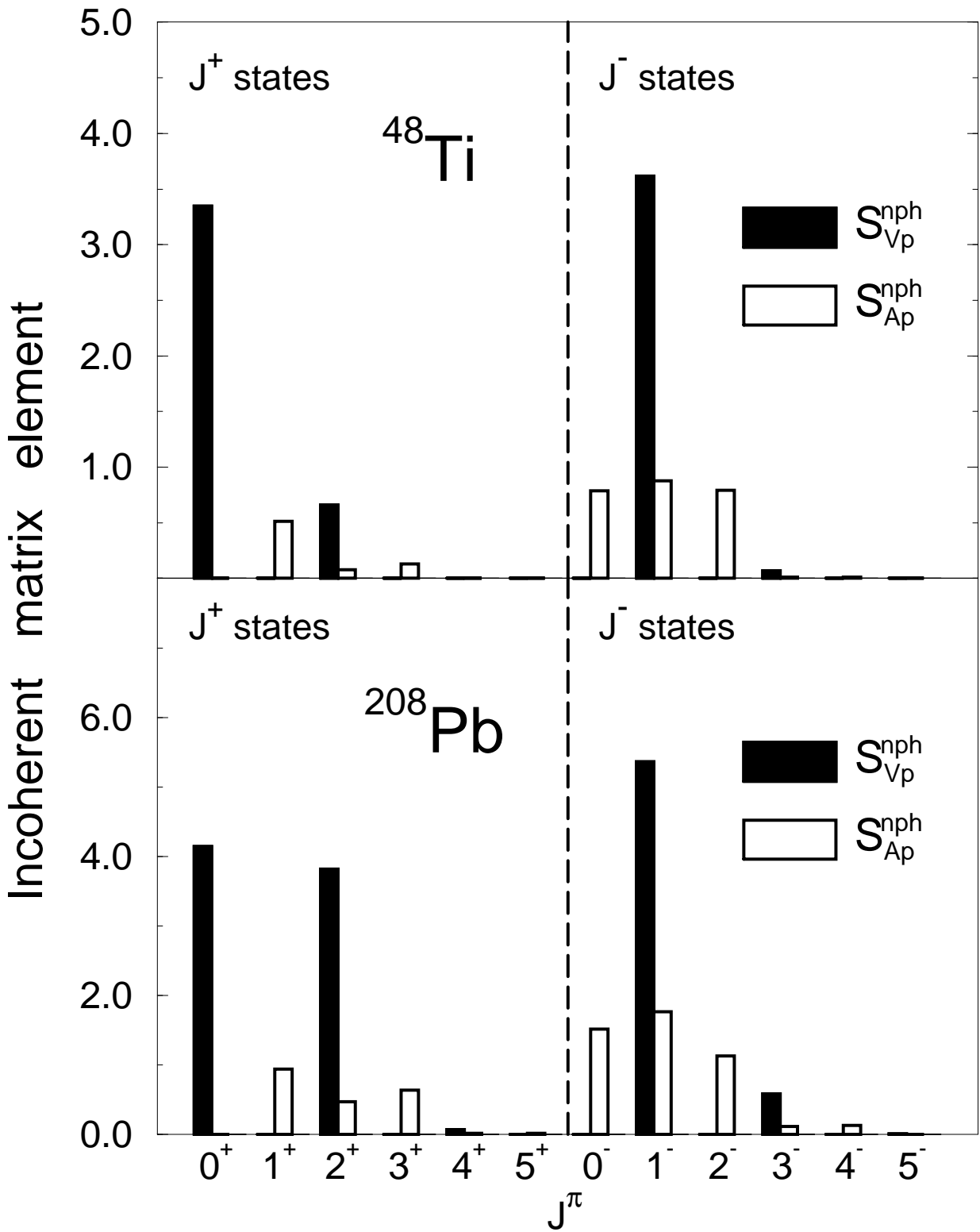
\end{document}